\documentclass[a4paper,10pt,twoside,UTF-8]{cpc-hepnp}

\usepackage{multicol}
\usepackage{graphicx}
\usepackage{subfig}
\usepackage{booktabs}
\usepackage{amssymb,bm,mathrsfs,bbm,amscd}
\usepackage[tbtags]{amsmath}
\usepackage{lastpage}
\usepackage{CJK}
\usepackage{algorithm}
\usepackage{algpseudocode}
\newcommand{\ucite}[1]{\textsuperscript{\cite{#1}}}

\makeatletter

\newenvironment{figurehere}
  {\def\@captype{figure}}
  {}
\makeatother
\begin{document}


\fancyhead[c]{\small Chinese Physics C~~~Vol. XX, No. X (201X)XXXXXX} \fancyfoot[C]{\small 010201-\thepage}

\footnotetext[0]{Received 14 March 2016}

\title{{The Study of Cosmic Ray Tomography
Using Multiple Scattering of Muons for Imaging of High-Z Materials
\thanks{Supported by National Natural Science Foundation of China under Grant Nos 11135002,and  11575268, and the foundation of Education Department of Hunan Province under Grant No 12B205.}}}

\author{%
 WANG Xiao-Dong$^{1}$
\quad Ye Kai-Xuan$^{2}$
\quad Li Yu-Lei$^{1}$
\quad Luo Wen$^{1}$
\quad Wu Hui-Yin$^{4}$
\quad Yang He-Run$^{3}$\\
\quad Chen Guo-Xiang$^{1}$
\quad Zhu Zhi-Chao$^{1}$
\quad Zhao Xiu-Liang $^{1;1}$\email{zhaoxiul@usc.edu.cn,wxdusc@yahoo.com}
}
\maketitle

\address{%
$^1$ {School of Nuclear Science and Technology, University of South China, Hengyang 421001, China}\\
$^2$ {Institute of Plasma Physics Chinese Academy of Sciences, University of Science and Technology of China, Hefei 230031, China }\\
$^3$ {Institute of Modern Physics, Chinese Academy of Sciences, Lanzhou, 730000, China}\\
$^4$ {School of Nuclear Science and Technology, Lanzhou University, Lanzhou 730000, China}\\
}

\begin{abstract}
Muon tomography is developing as a promising system to detect high-Z (atomic number) material for ensuring homeland security. In the present work, three kinds of spatial locations of materials which are made of aluminum, iron, lead and uranium are simulated with GEANT4 codes, which are horizontal, diagonal and vertical objects, respectively. Two statistical algorithms are used with MATLAB software to reconstruct the image of detected objects, which are the Point of Closet Approach (PoCA) and Maximum Likelihood Scattering-Expectation Maximization iterative algorithm (MLS-EM), respectively. Two analysis methods are used to evaluate the quality of reconstruction image, which are the Receiver Operating Characteristic (ROC) and the localization ROC (LROC) curves, respectively. The reconstructed results show that, compared with PoCA algorithm, MLS-EM can achieve a better image quality in both edge preserving and noise reduction. And according to the analysis of ROC (LROC) curves, it shows that MLS-EM algorithm can discriminate and exclude the presence and location of high-Z object with a high efficiency, which is more flexible with an different EM algorithm employed than prior work. Furthermore the MLS-EM iterative algorithm will be modified and ran in parallel executive way for improving the reconstruction speed.
\end{abstract}



\footnotetext[0]{\hspace*{-3mm}\raisebox{0.3ex}{$\scriptstyle\copyright$}2013
Chinese Physical Society and the Institute of High Energy Physics
of the Chinese Academy of Sciences and the Institute
of Modern Physics of the Chinese Academy of Sciences and IOP Publishing Ltd}%
\begin{multicols}{2}

\section{Introduction}
Cosmic-ray muons, coming from primary cosmic rays at deep space, with a limited flux around $1~muon/(min\cdot cm^2)$\ucite{PDBook} at sea level, can penetrate high-Z or dense material. Muon is scattering by the interaction of multiple Coulomb scattering (MCS) when it is close to the nuclei of atom. The scattering information of the object along the muon trajectory can be measured by a pair of position sensitive detectors such as Drift-Tube detectors\ucite{Schultz2003Cosmic,Bittner2016Development}, Gas Electron Multiplier (GEM)\ucite{Gnanvo2008Performance,Sturdy2016CMSexperiment}, Multi-gap Resistive Plate Chamber (MRPC)\ucite{Ye2008Study,Yu2015MAP} and so on, which are placed on both sides of the detected objects. By collecting the position information when muons are incoming and outgoing the sensitive area of detectors with a broad angular distribution of momentum, the 3D image of unknown object in container can be reconstructed.
\par
Muon tomography (MT) is considered as a promising technique using the natural particles in GeV with high penetrability to detect shielded packages. Comparing with traditional radiation ray tomography, such as X-ray and gamma ray radiography, the attractive features of muon tomography are no manufactured source, no artificial dose, and high sensitivity to special nuclear material (SNM)\ucite{Schultz2003Cosmic}. The perspective of MT is conducive to control the illegal transport of nuclear material and ensure homeland security .
\par
In this paper, a discrete tomographic reconstruction concept base on MCS and the feasibility of realizing the algorithm will be described firstly. Secondly, two reconstruction algorithms are accomplished, which are the Point of Closet Approach (PoCA) and a Maximum Likelihood Scattering-Expectation Maximization (MLS-EM). A different iterative Expectation Maximization (EM)\ucite{DEMPSTER1977Maximum,Green1990Bayesian} algorithm is introduced to find the ML estimate of scattering density profiles of material, which is efficient and flexible\ucite{Schultz2007Statistical} in medical image reconstruction. Lastly, the comparison results of the reconstruction image between the two algorithms are evaluated by three different simulated scenes of horizontal, diagonal and vertical objects using the GEANT4 Monte Carlo simulation codes. And the image quality are researched by the ROC (Receiver Operation Characteristic) curve as well as the localization ROC (LROC) using the MATLAB software.
\section{Concept of muon tomography}
\subsection{Tomography for MCS of muons}
The traditional tomography concept is illustrated in Fig.\ref{fig1}, where tomography refers to the reconstruction image of object from projections taken from many different directions. $M$ rays sample the object characteristic function when they pass through the imaging area along the line. When the $i^{th}$ ray passes the imaging area and is detected, its sampling (or signal) can be observed. The relationship between a ray's sampling and the discrete object characteristic function can be described as following ray-sum expression:
\begin{equation}\label{eq1}
P_i=\sum_j w_{ij} f_j,
\end{equation}
where, the weight $w_{ij}$ is the path length of the $i^{th}$ ray through the $j^{th}$ pixel or voxel. Solving the system of linear equations \eqref{eq1}, the reconstructed characteristic function $\bm{\hat{f}}$ can be estimated.
Since muon tomography is based on the traditional method, there are still several problems need to be modified:
\begin{itemize}
  \item The natural muons have a limited flux and come from the directions which have a broad angular distribution around zenith.
  \item The ray signal, namely, the multiple Coulomb scattering angle is stochastic with a zero-mean Gaussian, and the actual distribution even has heavier tails.
  \item The muon trajectories are not straight, therefore we may find a rough location and structure of the high-Z material.
\end{itemize}
\par
\begin{center}
\includegraphics[width=8cm,height=6cm]{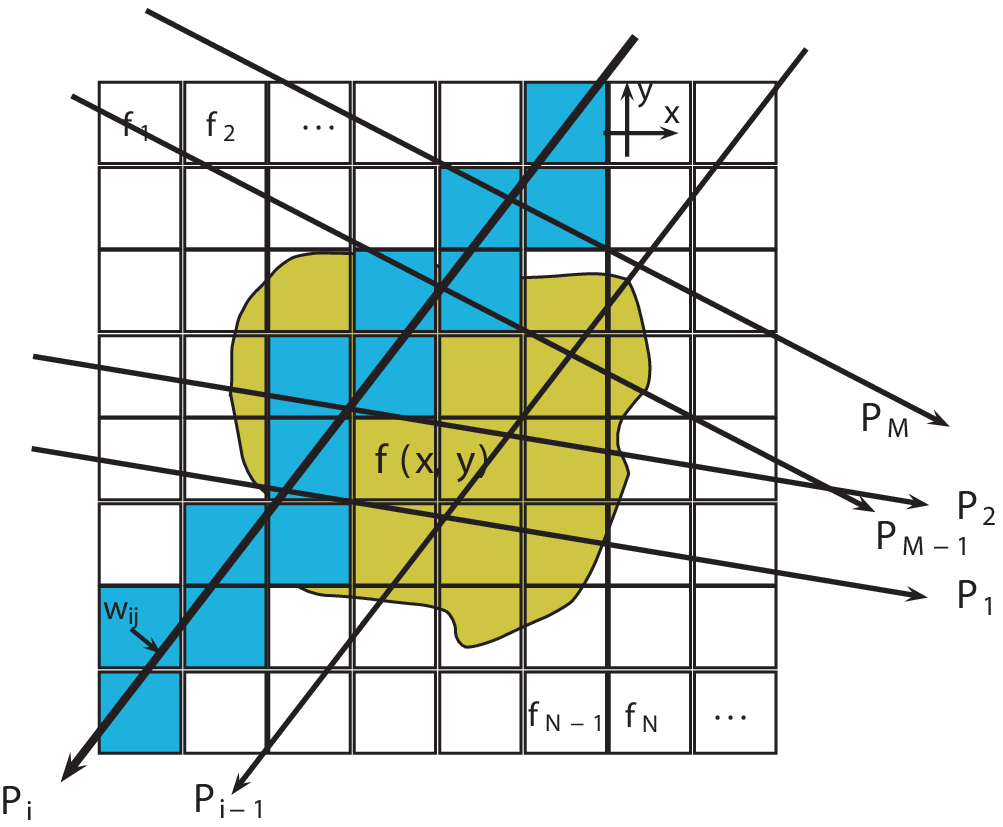}
\end{center}
\figcaption{\label{fig1} (Color online) The traditional tomography concept. A discrete model of the object characteristic function is adopted by assuming uniform values within each pixel or voxel, denoted by the values $f_1,f_2,\cdots,f_N$. And the rays sampling can be observed. }
As illustrated in Fig.\ref{fig2} where the scattering angle is exaggerated, the observed data $D_i$ of a muon is the scattering angle $\Delta\theta_i$. In muon tomography, we use the x and y planes for each of the scattering angle $\Delta\theta_i$ to add reconstructed information, $D_{x,i}~(D_{y,i})$ as:
\begin{equation}\label{eq2}
\begin{split}
D_{x,i}&=\Delta\theta_{x,i}=(\theta_{x,out}-\theta_{x,in})_i\\
D_{y,i}&=\Delta\theta_{y,i}=(\theta_{y,out}-\theta_{y,in})_i.
\end{split}
\end{equation}

The conditional probability distribution of the observed data $D_i$ may be approximated as a Gaussian\ucite{Schultz2007Statistical} with a zero mean and the variance $\Sigma_i$, given the scattering density distribution $\bm{\lambda}$, as:
\begin{equation}\label{eq3}
  P(D_{i}|\bm{\lambda})=\frac{1}{\sqrt{2\pi}|\Sigma_i|^{1/2}}exp{\left(-\frac{D_{i}^2}{2\Sigma_i}\right)}.
\end{equation}
\par
The variance $\Sigma_i$ can be express as
\begin{equation}\label{eq4}
  \Sigma_i =p^2_{r,i}\sum_{j}L_{ij}\lambda_j,
\end{equation}
where $L_{ij}$ is similar to $w_{ij}$, which is the path length of the $i^{th}$ ray through the $j^{th}$ voxel, and $p_{r,i}$ is the momentum ratio which is inversely proportional to $i^{th}$ muon momentum $p_{i}$.
\par
Considering the muon detector noise, the variance is refined as
\begin{equation}\label{eq5}
  \Sigma_i =C_i+p^2_{r,i}\sum_{j}L_{ij}\lambda_j,
\end{equation}
where, $C_i$ is the contribution of the detector noise\ucite{Schultz2003Cosmic}. The expression of \eqref{eq1}, \eqref{eq4} and \eqref{eq5} have a similar form, while \eqref{eq4} and \eqref{eq5} could express the relationship between the variance of scattering angle and the scattering density for the stochastic signal.
\begin{center}
\includegraphics[width=0.35\textwidth,keepaspectratio]{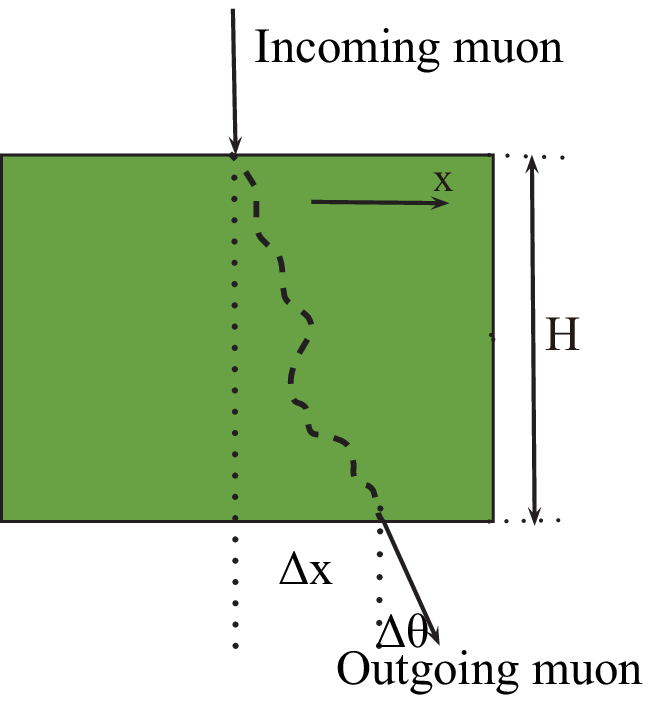}
\end{center}
\figcaption{\label{fig2} (Color online) A muon is scattering by the interaction of multiple Coulomb scattering passing through material, which can be described by the scattering angle $\Delta\theta$ and displacement $\Delta x$. Here, the path length of ray is approximately equal to
the thickness of material $H$.}
\subsection{Simulation platform by GEANT4}
The concept of muon tomography have be established above, then we need to use the Monte Carlo simulation package GEANT4 to simulate the MT spectrometer based on position sensitive detectors. The imaging area has a volume of $2\times2\times1~m^3$ filled with air, and sensitive are of $2\times2~m^2$.  To achieve the reconstruction result accordant with practical circumstances, the number of simulated muons is $2.0\times10^5$ corresponding to 5 min of exposure in experiment and the size of material cubes are $10\times10\times10~cm^3$. Using simulated data is important because that it can help us understand muon behavior and provide us "true" information of each event which in the experiment may be not obtained.
 \par
\begin{center}
\figurehere
\includegraphics[width=0.5\textwidth,keepaspectratio]{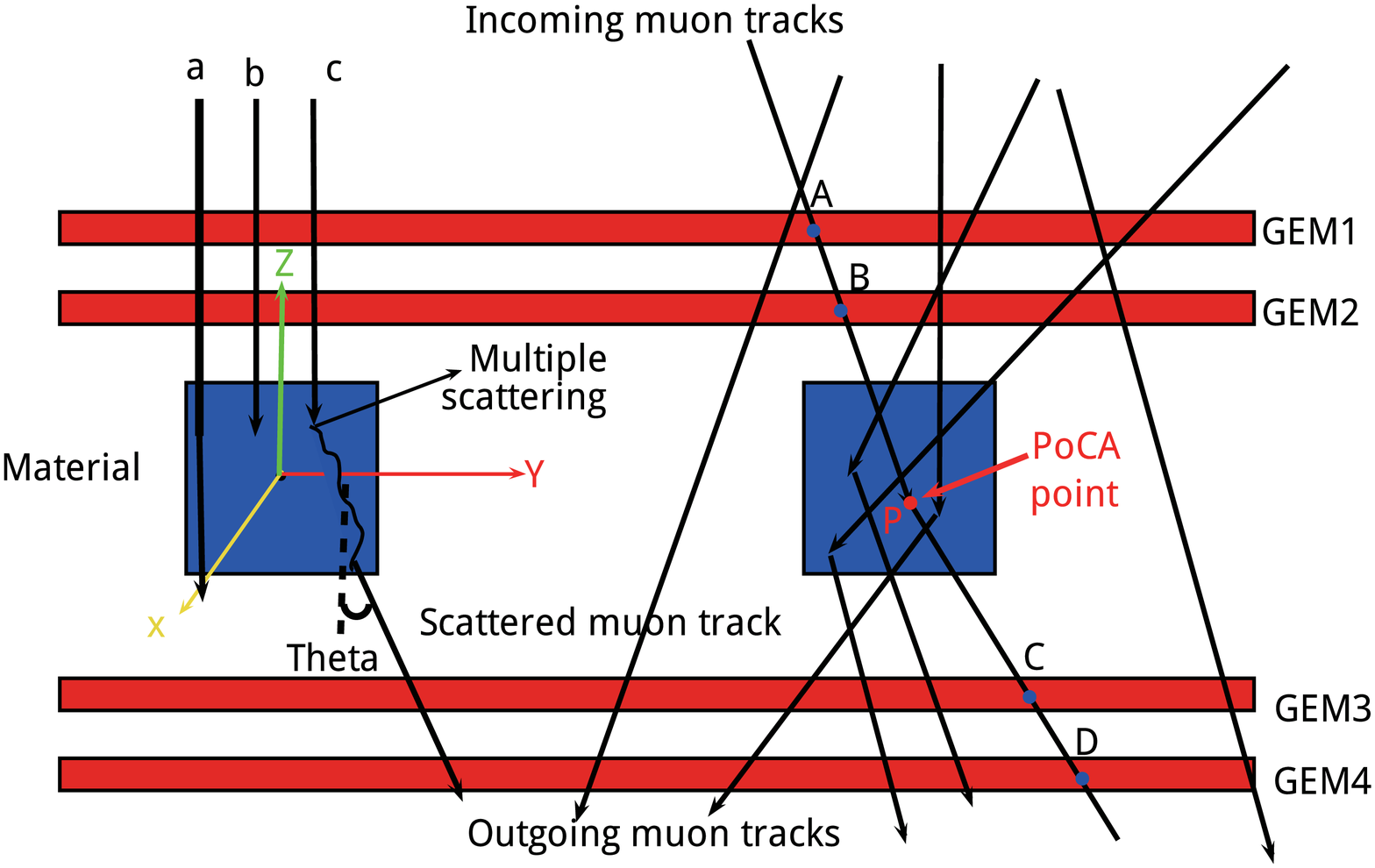}
\end{center}
\figcaption{(Color online) The muon tomography concept. The muon interactions, a, b and c, are shown separately for express clearly: transmission, stop and MCS. The incident points A, B and the exiting points C,D are the positions of a incoming and outgoing muon, respectively, and other information is shown in the figure.\label{t6} }
\par
Fig.\ref{t6} shows that a group of position sensitive detectors are located above the object, and one below. These detectors are "perfect" which can record the measurements such as positions, angles and momentums of incoming and outgoing muons, respectively. Besides, muons pass through the volume and their scattering manners depend on the atomic number, Z, and the material density.
\par
\par
The physics list QGSP-BERT-HP in GEANT4 is used to completely describe the physical process of muon, as a, b, c shown in Fig.\ref{t6}. The spectrum of generated muon is established by the spectrum of cosmic ray muons in the range of 3 to 100~GeV, which obeys the empirical formula\ucite{Nakamura2010}:
\begin{equation}\label{eq6}
  \frac{dI}{dE\times dcos\theta}=0.14E^{-0.27}\left( \frac{1}{1+\displaystyle\frac{1.1E cos\theta}{115 GeV}}+\frac{0.54}{1+\displaystyle\frac{1.1E cos\theta}{850 GeV}} \right),
\end{equation}
where, $\theta$ is plane angle from vertical and $E$ is the muon energy.

\section{Reconstruction algorithms}
\subsection{PoCA reconstruction algorithm}
Fig.\ref{thetas} shows that the probability density distribution of scattering angle $\Delta\theta$ can be approximated as Gaussian\ucite{Nakamura2010} distribution (as the red fitting curve shown) when muons are passing through the aluminum, iron, lead and uranium cubes. The variance of scattering angle is a function of atomic Z number, and the results show that the RMS scattering (mrad) are about 4.33 for aluminum, 10.04 for iron, 18.69 for lead, and 25.24 for uranium, respectively.

The PoCA is a simple geometric algorithm under the assumption of single scattering in each event. This algorithm can quickly discriminate the location and structure of object, and be sensitive to high-Z material, which has been validated by many famous labs in experiment\ucite{Borozdin2003Surveillance,Gnanvo2008Performance,Bogolyubskiy2008First,Cox2008Detector}.
\par
Firstly, the scattering angle is around of milliradians, the track of each muon can be computed as a straight line connecting the incoming and outgoing points which are detected by above and below detectors. Secondly, it should be assumed that the scatter of each event only occurred once time on the closest point to the incoming and outgoing tracks, which is called as the PoCA point. In three dimensions space, the incident and scattered tracks may not be coplanar and not intersect at a point. The point closet to the each pair line (incoming and outgoing) are computed by solving a linear algebraic formulation and the midpoint of common perpendicular are taken as the PoCA point. Fig.\ref{t8} are the schematic of PoCA algorithm in the voxellation of imaging area.

\begin{center}
\figurehere
\includegraphics[width=0.5\textwidth,keepaspectratio]{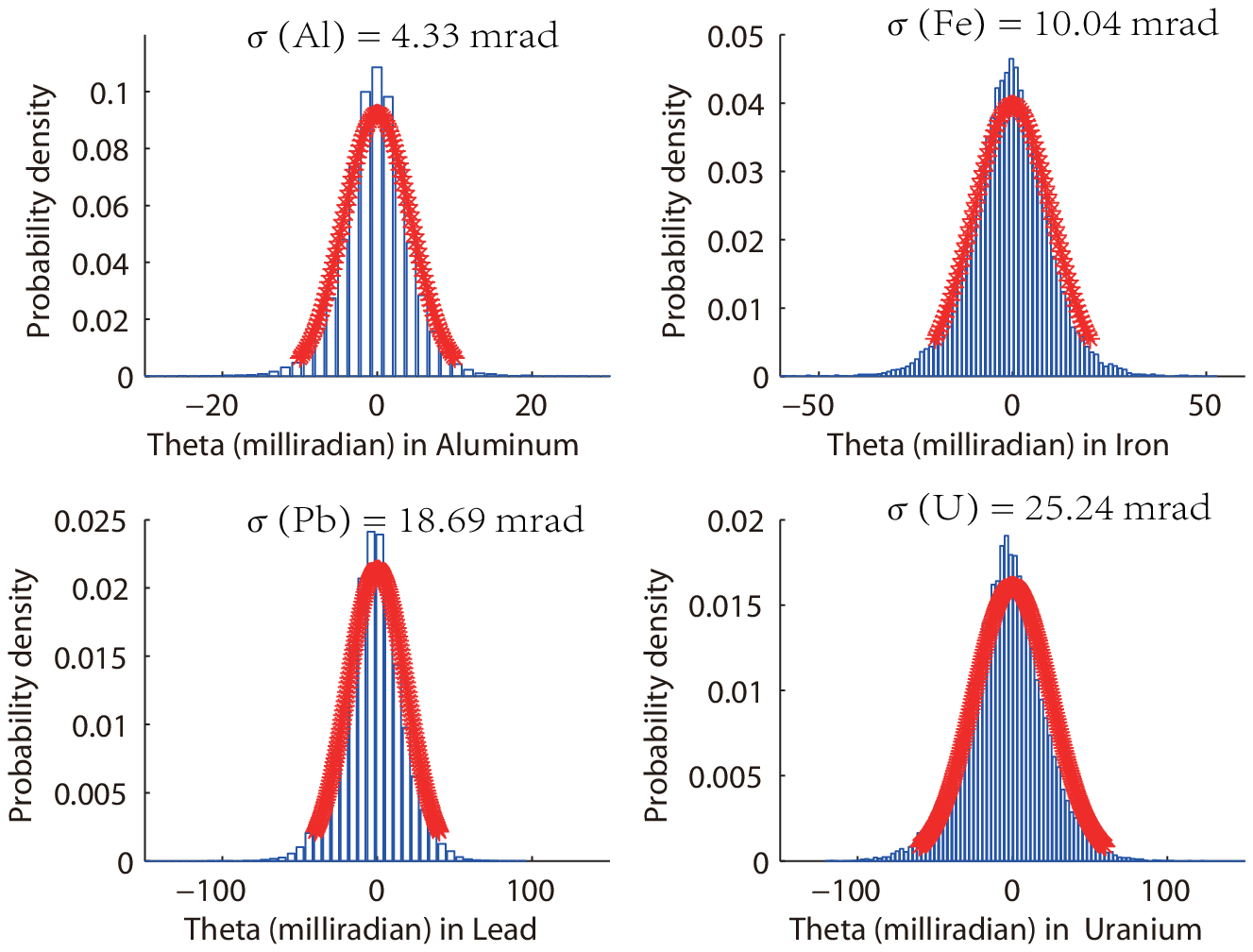}
\end{center}
\figcaption{\label{thetas} (Color online) The probability density distribution of scattering angle may be approximated as a zero-mean Gaussian. The RMS scattering are about 4.33 (mrad) for Al (upper-left), 10.04 for Fe (upper-right), 18.69 for Pb (lower-left), and 25.24 for U (lower-right). The blue histogram and red curve are the simulated and fitting results, respectively.}
\par
\begin{center}
\figurehere
\includegraphics[width=0.5\textwidth,keepaspectratio]{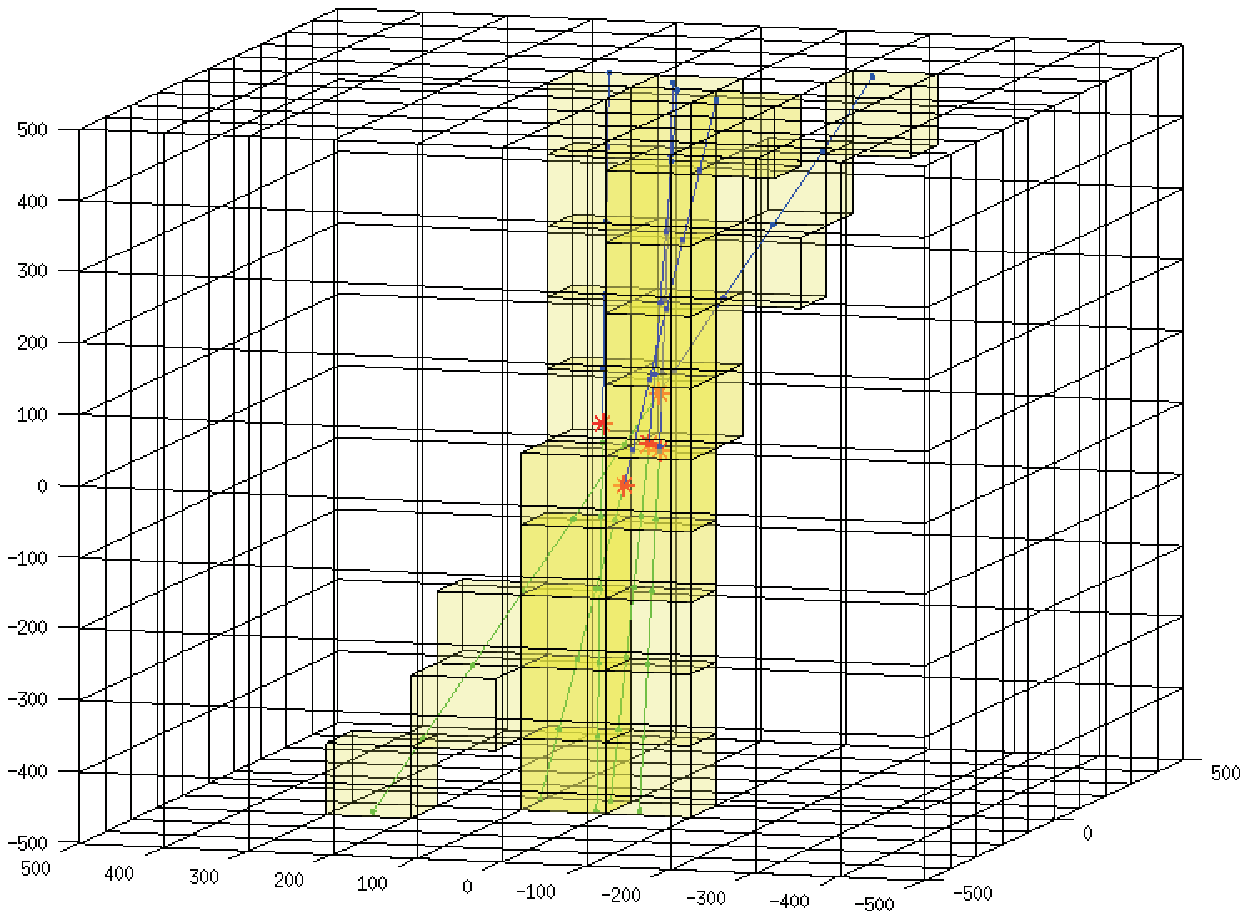}
\end{center}
\figcaption{(Color online) The PoCA algorithm schematic. The blue and green lines are incident and exiting tracks respectively. The yellow cubes are voxels through which the muons pass, and the red points are assumed to be PoCA points. \label{t8} }
\par
Thirdly, the ray signal can be marked as $s_i$ and defined as:
\begin{equation}\label{eq7}
\begin{split}
  s_i&=\frac{\sqrt{(\Delta\theta^2_{x,i}+\Delta\theta^2_{y,i})}}{2}\\
     &=\left[\frac{(\theta_{x,out}-\theta_{x,in})^2+(\theta_{y,out}-\theta_{y,in})^2}{2}\right]_i.
\end{split}
\end{equation}
\par
Moreover, since muons come from different directions so that path length is different for each muon, we modify \eqref{eq4} to compute the scattering density estimate $\bm{\hat{\lambda}}$ for reconstruction.
\begin{equation}\label{eq8}
  \lambda_j=\frac{1}{p^2_r}\frac{\sigma^2_\theta}{L}=\sum_{i:L_{ij}\neq 0}\frac{s_{ij}^2}{p^2_{r,i}M_jL_{ij}},
\end{equation}
where, the ray signal $s_{ij}$ is:
\begin{equation}\label{eq9}
  s_{ij} = \left\{\begin{array}{rcl}
         s_i& Poca~voxel\\
         0  & along~path~except~Poca~voxel
         \end{array}
         \right
         .
\end{equation}
\par
Finally, the corresponding flow diagrams of PoCA algorithm are shown in Tab.\ref{alg:PoCA}, and the complexity is o(M) for M rays.
\par

\begin{algorithm}[H] 
\tabcaption{\label{alg:PoCA}The summary of the PoCA algorithm for muon tomography}
\begin{algorithmic}[1]                
\Require                           
    Measurement data of track angle, position, and momentum of each muon, $(\theta_x,\theta_y,x,y,z)_{in}$ and $(\theta_x,\theta_y,x,y,z,p^2_{r})_{out}$.
\Ensure                            
    Estimate of scattering density, $\lambda_{j,poca}$.
\For{$j=1:N$}
\For{$i=1:M$}
\State
    Compute the path length of each muon with each voxel, $L_{ij}$.
\State Tag the index of voxel through which the muons pass, $V_{num}$.
\State Tag the number of PoCA voxel, $Vpoca_{num}$.
\State Compute the ray signal of each muon with each voxel, $s_{ij}$ using \eqref{eq9}.
\EndFor
\State Calculate the number of muons hitting the $j^{th}$ voxel, $M_j$£»
\State Estimate $\lambda_{j}=\sum_{i:L_{ij}\neq 0}\displaystyle\frac{s_{ij}^2}{p^2_{r,i}M_jL_{ij}}$, using \eqref{eq8}.
\EndFor
\State
\Return $\lambda_{j,poca}=\lambda_j$£»               
\end{algorithmic}
\end{algorithm}

\par

\subsection{MLS-EM reconstruction algorithm}
A better algorithm, Maximun Likelihood Scattering-Expectation Maximization (MLS-EM), is accomplished further by using information of scattering angles, which distribute the scattering location along the ray track instead of assigning to the PoCA point according to  probability statistics. In order to compute the ray path lengths through voxels, the entry points to PoCA points to exit points are connected to estimate.
\par
The total function of muon data may be written as:
\begin{equation}\label{eq10}
P(\bm{D}|\bm{\lambda})=\prod_{i}P(D_i|\bm{\lambda}),
\end{equation}
where the conditional probability distribution $P(D_i|\bm{\lambda})$ is given by \eqref{eq3}.
\par
The MLS estimation of the scattering density $\hat{\bm{\lambda}}$ can be solved by maximizing the log likelihood function:
\begin{equation}
\begin{split}\label{eq11}
  \hat{\bm{\lambda}}_{MLS}&=arg\max_{\bm{\lambda}>\bm{\lambda}_{Air}}LP(\bm{\lambda|D})\\
                     &=arg\max_{\bm{\lambda}>\bm{\lambda}_{Air}}\left(\frac{1}{2}\sum_i\left(-log(\Sigma_i)-\frac{D_i^2}{\Sigma_i}\right)\right),
\end{split}
\end{equation}
an EM algorithm to maximize the log likelihood function are developed
to calculating the scattering density estimate, which is more efficient and flexible than traditional method.
\par
The following MLS-EM update equation for the scattering density at the $n^{th}$ iteration are derived as:
\begin{equation}\label{eq12}
  \hat{\lambda}_{j,MLS-EM}^{n+1}=\frac{1}{2}median_{i:L_{ij\neq0}}B^n_{ij}.
\end{equation}
\par
Since the measurements in x and y are independent, $B_{ij}$ can be computes as the average of $B_{x,ij}$ and $B_{y,ij}$:
\begin{equation}\label{eq13}
  B^n_{x,ij}(B^n_{y,ij})=2\lambda_j^n+\left(\frac{D^2_{x,i}(D^2_{y,i})L_{ij}}{\Sigma_i^2}-\frac{L_{ij}}{\Sigma_i}\right)\times p_{r,i}^2(\lambda_j^n)^2.
\end{equation}
\par
More detailed information can be found in Ref.\ucite{Schultz2007Statistical}, and the executing process of MLS-EM algorithm is shown in Tab.\ref{alg:MLS-EM}. The asymptotic time complexity of the MLS-EM algorithm is o(MNI), where M is the number of muons, N is the number of voxels in imaging area, and I is the number of iterations. Besides, the memory footprint is o(M+N).
\par
\begin{algorithm}[H]         
\tabcaption{\label{alg:MLS-EM}The summary of the MLS-EM algorithm for muon tomography}
\begin{algorithmic}[1]                
\Require                           
    Measurement data of the scattering angle and momentum of each muon, $(\Delta\theta_x,\Delta\theta_y,p^2_r)_i$.
\Ensure                            
    Estimate of scattering density, $\lambda_{j,mls-em}$.
\State
    Compute the path length for each muon voxel pair, $L_{ij}~(1\leq j\leq N \& 1\leq i\leq M)$.
\State
    Set the initial value of each voxel with the scattering density of air, $\lambda_{j,old}=\lambda_{air}$.
\For{$iter\leq k $}
    \State Compute the variance of each muon, $\Sigma_i$ using \eqref{eq5}.
    \State Compute the update terms, $B^n_{ij}$ using the \eqref{eq13}.
    \State Estimate the scattering density of each voxel at $n^{th}$ iteration, $\lambda_{j,new}=\lambda^{n+1}_{j}$ using \eqref{eq12}.
    \State Update the scattering density, $\lambda_{j,old}=\lambda_{j,new}$.
    \State Update the iteration, $iter=iter+1$.
\EndFor
\State
\Return $\lambda_{j,mls-em}=\lambda_{j,new}$£»               
\end{algorithmic}
\end{algorithm}
\par
\section{Results and Analysis}
\subsection{Results}
\begin{figure*}[h]
\begin{center}
\includegraphics[width=1.1\textwidth,keepaspectratio]{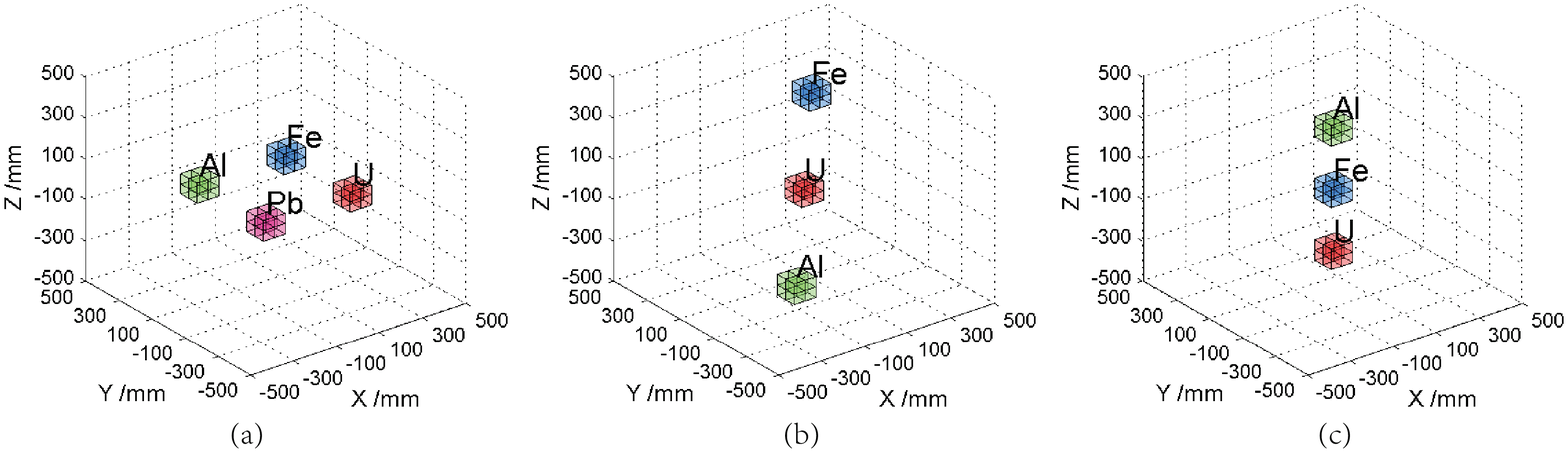}
\caption{(Color online) The perspective view of the simulated horizontal (a), diagonal (b) and vertical (c) scenes.\label{thor} }
\end{center}
\end{figure*}

Fig.\ref{thor} shows the perspective view of three different scenes for comparison between PoCA and MLS-EM algorithms: the horizontal, diagonal and vertical objects. The number of incident muons  is $2.0\times10^5$, corresponding to about 5 mins of exposure in experiment. The voxels are the size of $50\times50\times50~mm^3$ placed in the imaging area whose size is $2\times2\times1~m^3$.
\par
\begin{figure*}[h]
\begin{center}
\includegraphics[width=\textwidth,keepaspectratio]{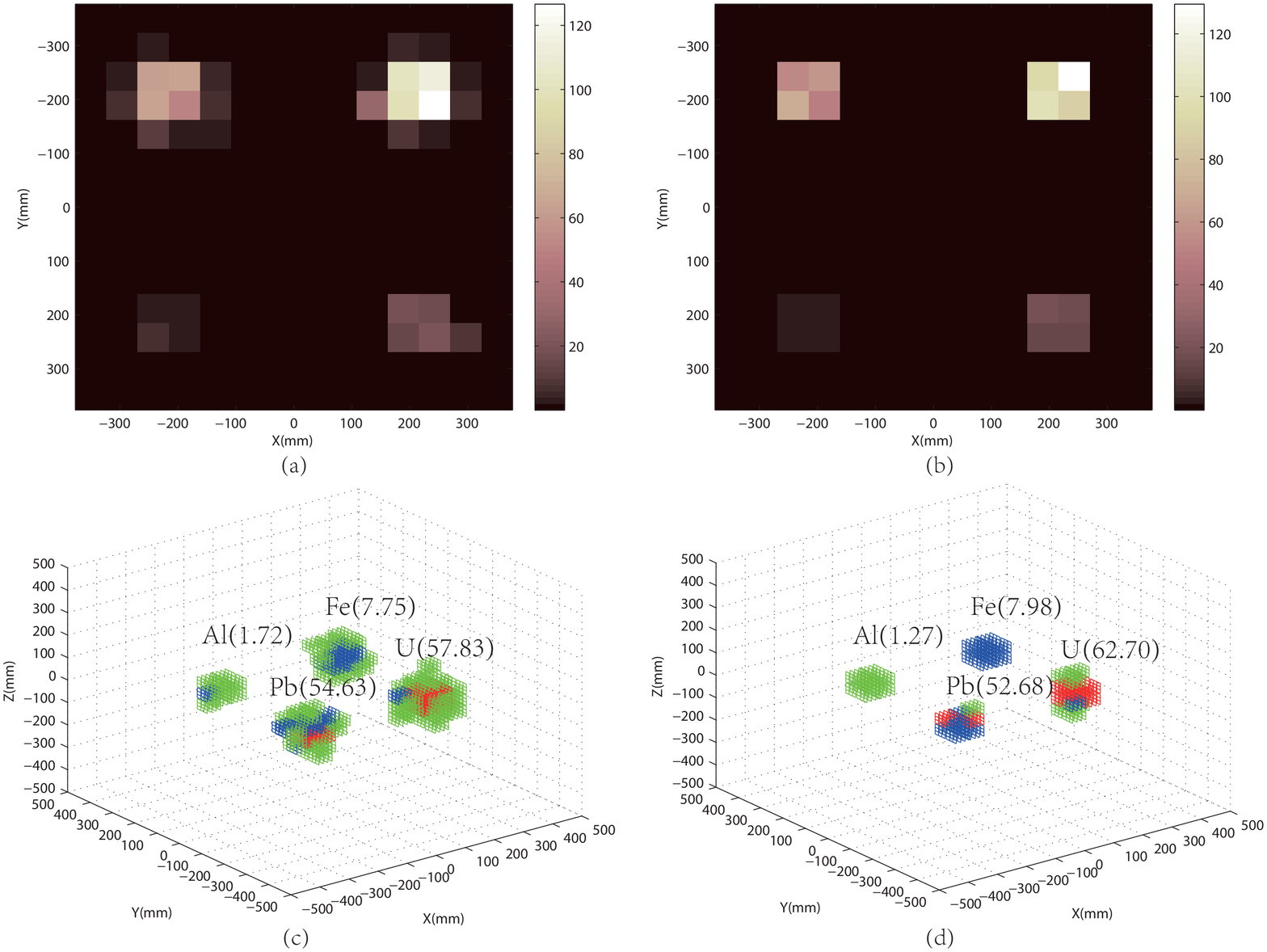}
\caption{ (Color online) The comparison reconstruction in 2D (a,b) and 3D (c,d) of the horizontal material between PoCA (a,c) algorithm with MLS-EM (b,d) algorithm.\label{tmls-1} }
\end{center}
\end{figure*}

Fig.\ref{tmls-1} are the comparison results of horizontal scene between PoCA and MLS-EM. In the reconstructed image, the voxels with scattering density in a range of [-0.5,5) are colored by green (the low-Z), [5,30) blue (the medium-Z), and [30,$+\inf.$] red (the high-Z), respectively. In the MLS-EM reconstruction, the maximum times of iteration is set to 100, and the initial value of voxel scattering density is set to be of air: $\bm{\lambda}^0=10^{-6}$\ucite{Yu2015MAP}. Furthermore, the averages of scattering density in the voxels where objects appear are used to demonstrate the results in brackets. Fig.\ref{tmls-1}(a,c) are the 2D and 3D reconstructed images using PoCA algorithm, respectively, which shows that the scattering density estimate of (Al, Fe, Pb, U) are (1.72, 7.75, 54.63, 57.83 $mrad^2/cm$). Fig.\ref{tmls-1}(b,d) are the 2D and 3D reconstructed images using MLS-EM algorithm, respectively, which shows that the scattering density estimate of (Al, Fe, Pb, U) are (1.27, 7.98, 52.68, 62.70).
\par
\begin{figure*}[h]
\begin{center}
\includegraphics[width=\textwidth,keepaspectratio]{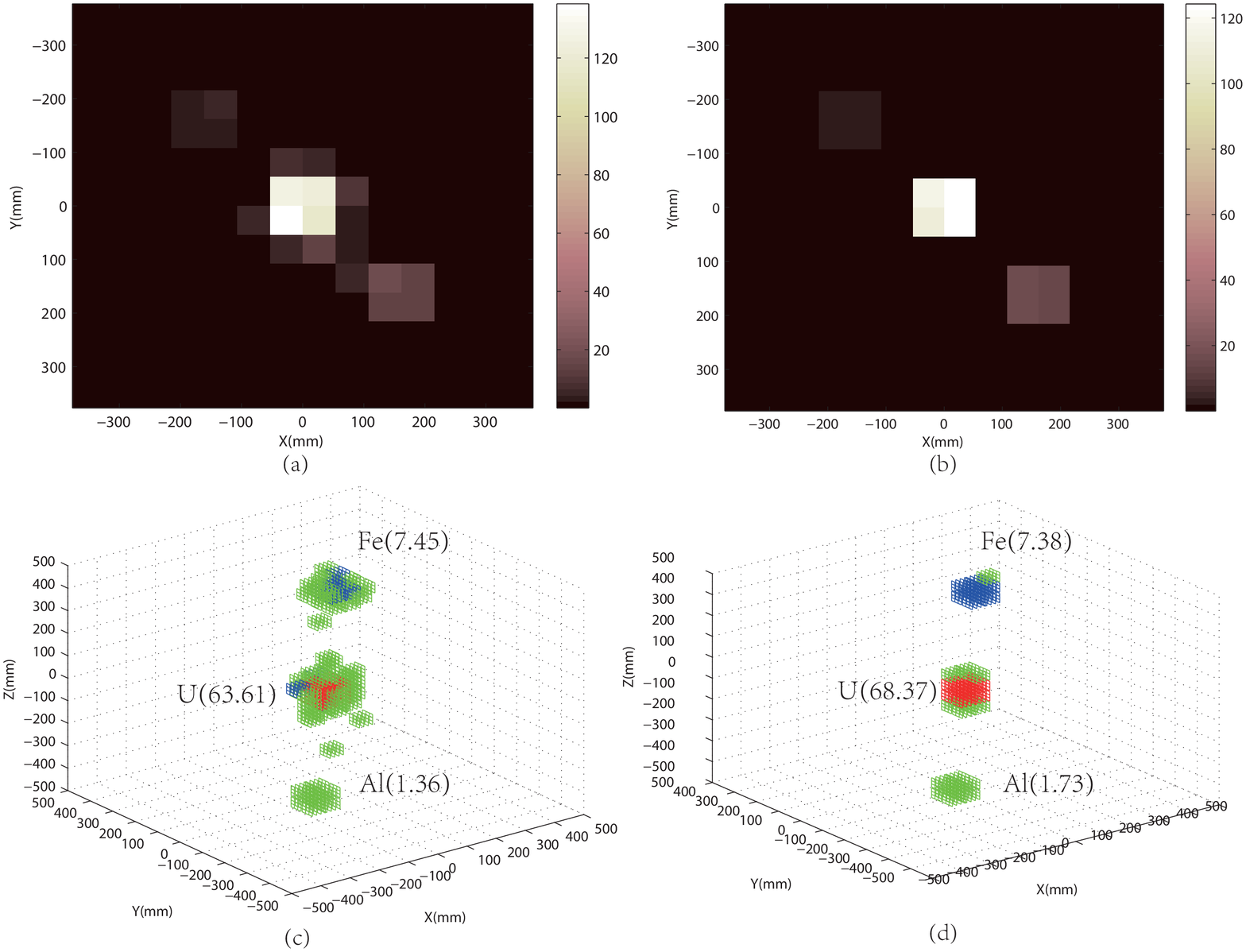}
\caption{ (Color online) The comparison reconstruction in 2D (a,b) and 3D (c,d) of the diagonal material between PoCA (a,c) algorithm with MLS-EM (b,d) algorithm.\label{tmls-2} }
\end{center}
\end{figure*}
\begin{figure*}[h]
\begin{center}
\includegraphics[width=\textwidth,keepaspectratio]{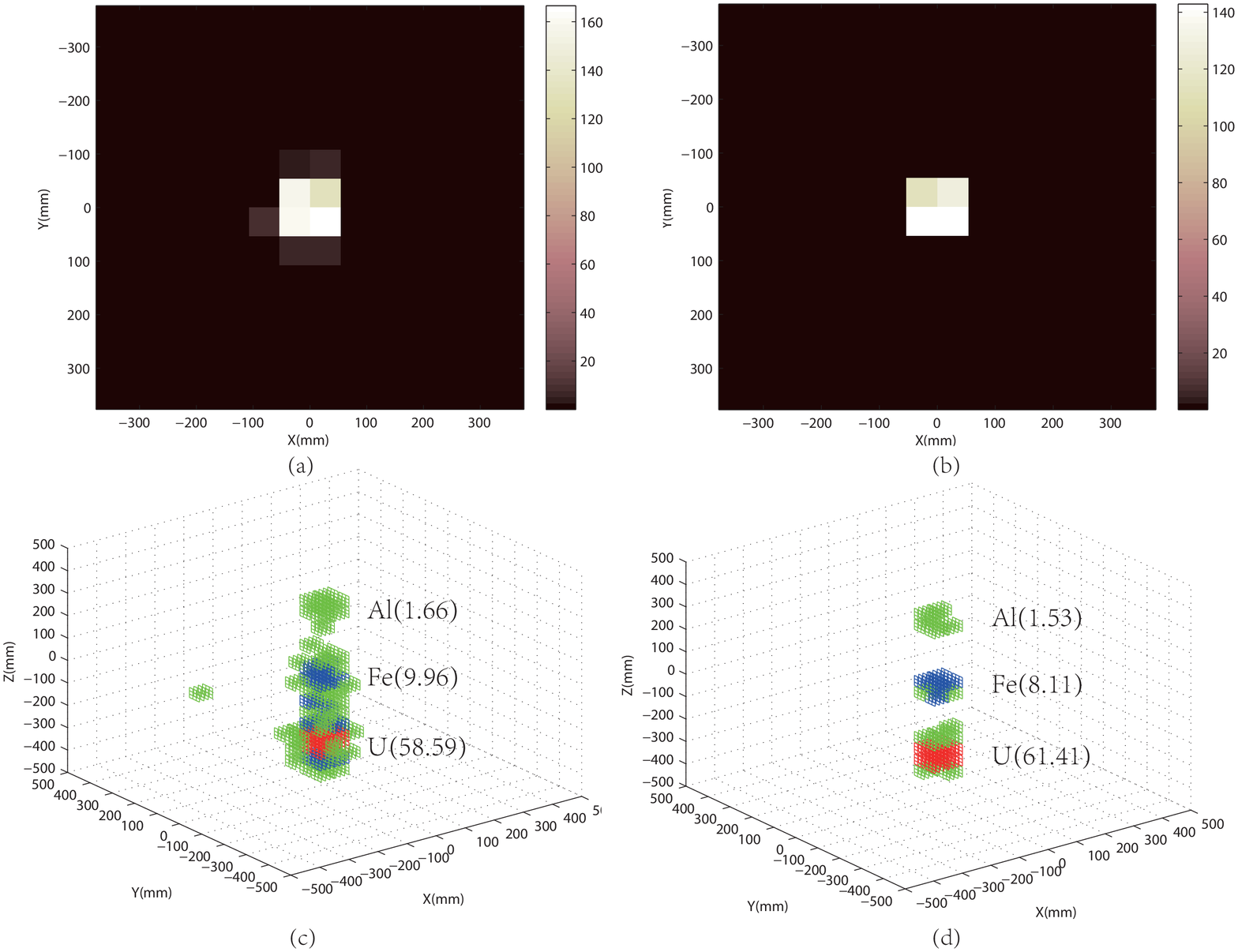}
\caption{ (Color online) The comparison reconstruction in 2D (a,b) and 3D (c,d) of the vertical material between PoCA (a,c) algorithm with MLS-EM (b,d) algorithm.\label{tmls-3} }
\end{center}
\end{figure*}

Fig.\ref{tmls-2} display the comparison results of diagonal scene between PoCA and MLS-EM, respectively, which show that the scattering density of (Al, Fe, U) are (1.36, 7.45, 63.61) using PoCA reconstruction and (1.73, 7.38, 68.37) using MLS-EM correspondingly. Fig.\ref{tmls-3} are the comparison results of vertical scene between PoCA and MLS-EM, respectively, which reflect that the scattering density of (Al, Fe, U) are (1.66, 9.96, 58.59) using PoCA reconstruction and (1.53, 8.11, 61.41) using MLS-EM correspondingly. Compared with the PoCA algorithm, the advantage of the MLS-EM algorithm is that it can reconstruct better location and appearance, particularly in vertical scenes, while the disadvantage of the MLS-EM algorithm is the complexity and much more consumptive of memory. The reconstructed time with PoCA algorithm is tens of seconds, but the time with MLS-EM is 1000-2000 s relatively.
\section{Analysis }
The goal of muon tomography is to discriminate and exclude the presence of high-Z material which can achieve a high efficiency and keeping false positives rate low in a short reconstructed time. We plot the ROC curve and the localization ROC (LROC) curve that has been commonly  used in the binary discrimination system to evaluate the image quality between PoCA and MLS-EM reconstruction.
\par
\begin{center}
\includegraphics[width=0.5\textwidth,height=0.35\textwidth]{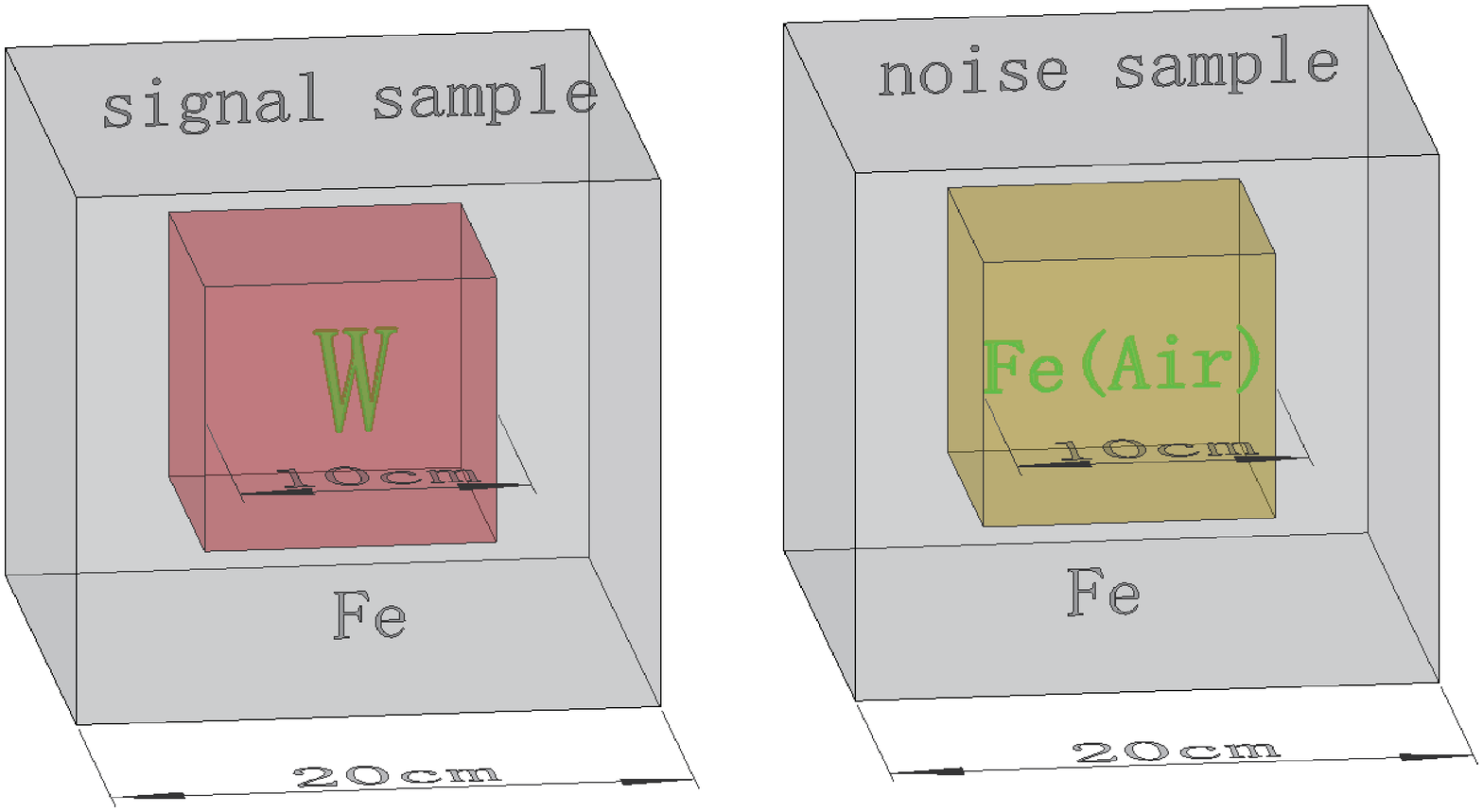}
\end{center}
\figcaption{(Color online) The ROC (LROC) analysis illustration for muon tomography.\label{troc} }
\par
Three sets of samples are generated as Fig.\ref{troc} shown. A set of sample hiding the target (a tungsten cube inside the iron volume) is regarded as the "signal" sample, the other two sets of sample without the target (empty or iron cube inside the volume) are regarded as the "noise" samples. The 50 images per set are used to test when the event of muon is $8.0\times10^4$ or less. The performance statistic is the maximum of scattering density , $\lambda_{max}$, in the image which is compared with a selected threshold, $T$. The true positive rate (TPR, sensitivity) is considered as the probability to trigger the alert in the "signal" image when $\lambda_{max}>T$. Otherwise, the false positive rate (FPR, 1-specificity) is defined as the probability to exclude the presence of target in the "noise" image when $\lambda_{max}<T$. Varying the threshold can generate a series of pair of sensitivity and specificity. The sensitivity plotted against 1-specificity is the ROC curve, and the perfect method would yield a point in the upper-left corner of the ROC curve. In that case, the area under the curve (AUC) is equal to 1.
\par
\begin{figure*}[!htbp]
\begin{center}
\includegraphics[width=\textwidth,keepaspectratio]{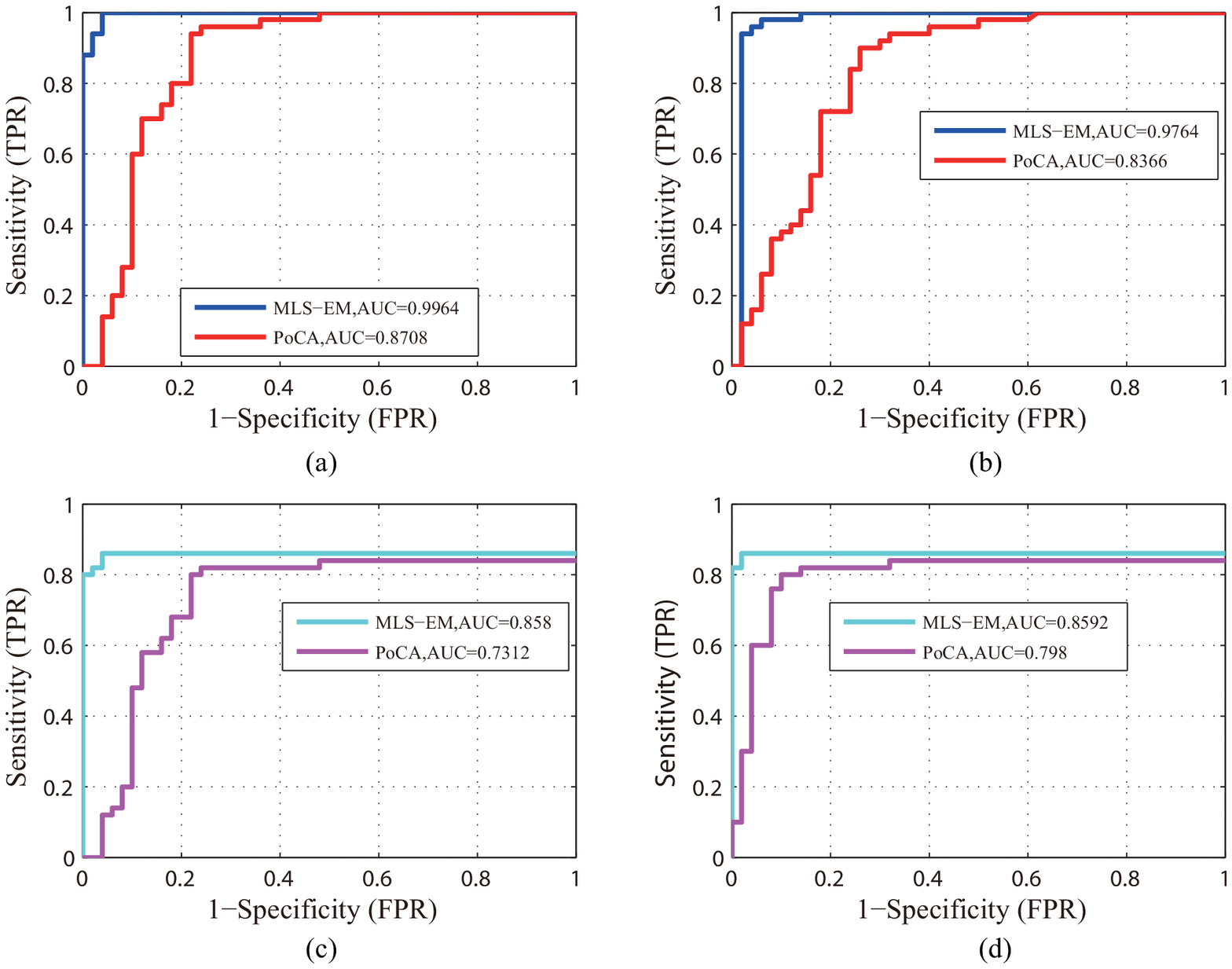}
\end{center}
\caption{ (Color online) The analysis of ROC (LROC) and AUC between PoCA and MLS-EM reconstruction: (a) ROC: W vs. Air, (b) ROC: W vs. Fe, (c) LROC: W vs. Air, (d) LROC: W vs. Fe.\label{troc-result} }
\end{figure*}

To consider the localization, the LROC curves are also performed, which have been used in the medical imaging community for assessing the lesion localization performance\ucite{Wang2009Bayesian}. The test statistic in LROC differ in that the true positive appears as if and only if the $\lambda_{max}$ is obtained at the target location in the "signal" sample. Fig.\ref{troc-result} exhibits the analysis of ROC and LROC curve between PoCA and MLS-EM algorithm. It is clear that no matter at any given PFR, MLS-EM can achieve much higher TPR than PoCA algorithm. For discriminating the signal sample from the noise sample, the MLS-EM reconstruction achieve AUC of 0.9964 in the air noise, 0.9764 in the iron noise for the ROC curve, 0.858 in the air noise, and 0.8592 in the iron noise for the LROC curve, respectively. The results are much higher than those of PoCA reconstruction (0.8708, 0.8366, 0.7312, and 0.798 correspondingly). The comparison reflects that the MLS-EM algorithm can significantly increase the performance of ROC and LROC to the PoCA approach.
\par

\section{Conclusion}
This paper describes two tomographic reconstructions based on MCS of natural muons. The instance of EM method is employed to solve large size ML problems which differs and more efficient. The simulated results and analysis suggest that, with a reasonable exposure time, PoCA algorithm is a qualitative algorithm, and compared with PoCA, MLS-EM algorithm can significantly improve the imaging performance of muon tomography.
\par
Considering the computation and memory usage of a standard PC and each voxel updated separately, speeding up the iterative reconstruction by transforming to parallel implementation will be studied in the further work. Besides, a construction of a small-sized MT based on GEM detectors is also in progress for applying these algorithms to experimental data to improve the algorithm results.

\acknowledgments{}
\end{multicols}

\vspace{10mm}

\begin{multicols}{2}

\end{multicols}

\vspace{-1mm}
\centerline{\rule{80mm}{0.1pt}}
\vspace{2mm}

\begin{multicols}{2}

\end{multicols}

\clearpage
\end{document}